# Improving the security of multiparty quantum secret sharing based on entanglement swapping against participant attack

Song Lin[a,b,*], Fei Gao[a], Fen-Zhuo Guo[a], Qiao-Yan Wen[a], Fu-Chen Zhu[c]

a *School of Science, Beijing University of Posts and Telecommunications, Beijing 100876, China*
b *School of mathematics and computer science, Fujian Normal University, Fuzhou 350007, China*
c *National Laboratory for Modern Communications, PO Box 810, Chengdu 610041, China*

In a recent paper [Z. J. Zhang and Z. X. Man, Phys. Rev. A **72**, 022303(2005)], a multiparty quantum secret sharing protocol based on entanglement swapping was presented. However, as we show, this protocol is insecure in the sense that an unauthorized agent group can recover the secret from the dealer. Hence, we propose an improved version of this protocol which can stand against this kind of attack.

With the development of quantum technology, quantum cryptography has become a hot research topic in the field of information security. As one important branch of quantum cryptography, quantum secret sharing (QSS) has attracted much attention [1-6]. In a simplest secret sharing, Alice, the dealer, wants that a secret is shared between her two agents (i.e., Bob and Charlie) so that it can be recovered only when they collaborate. Equivalently, only one agent, say Bob, cannot obtain any information about the secret. That is to say, Bob and Charlie constitute an authorized agent group, of which anyone, Bob or Charlie, is unauthorized. In general, during a multiparty quantum secret sharing (MQSS) process, there are more than two agents and not all of them are honest. Therefore, a secure QSS protocol should be able to ensure that any unauthorized agent group cannot elicit information about the secret. However, when discussing the security of a MQSS protocol, people tend to ignore the attack from the real agents. As mentioned in Ref. [7], a participant generally has more power to attack than an outside eavesdropper. So, we should pay more attention to the participant attack in the procedure of designing a secure MQSS protocol.

In a recent paper[6], Zhang et al proposed a multiparty quantum secret sharing protocol based on entanglement swapping. This protocol has several good features due to using Bell state and dense code. For example, it is easy to be implemented, and it achieves a high efficiency. But it is a pity that this protocol has a drawback. That is, some unauthorized agent groups may recover the secret by utilizing a special strategy. Consequently, this protocol is insecure against the participant attack.

Let us start with the brief description of MQSS protocol presented in Ref. [6], which we will call Zhang-Man protocol later. Without loss of generality, we take the four-party QSS protocol as our example. In such a protocol, Alice will split her key (the secret) into three pieces and then distribute them to her agents Bob, Charlie, and David, respectively. The three agents can deduce Alice's key if and only if they cooperate. The particular procedure is as follows (see Fig. 1). Firstly, Alice, Bob, Charlie, and David prepare an EPR pair $|\Psi^-\rangle_{12}$, $|\Psi^-\rangle_{34}$, $|\Psi^-\rangle_{56}$, and $|\Psi^-\rangle_{78}$ respectively, where

$$|\Psi^-\rangle = \frac{1}{\sqrt{2}}(|0\rangle|1\rangle - |1\rangle|0\rangle)$$

[see Fig.1 (a)]. Secondly, each of them sends a qubit from his/her EPR pair to the next person [see Fig.1 (b)]. With certain probability, Alice chooses the detecting mode where all the users check whether the qubits are transmitted in a secure manner (here the particular process to detect eavesdropping is not important to us, so we do not describe it in detail). Otherwise, in the message mode, Alice performs randomly one of the following four local unitary operations $\{u_1, u_2, u_3, u_4\}$ on qubit 1. Here $u_1 = |0\rangle\langle 0| + |1\rangle\langle 1|$, $u_2 = |0\rangle\langle 0| - |1\rangle\langle 1|$, $u_3 = |1\rangle\langle 0| + |0\rangle\langle 1|$ and $u_4 = |0\rangle\langle 1| - |1\rangle\langle 0|$. These operations represent Alice's secret, {"00","01","10","11"}, respectively. Afterwards, Alice performs a Bell-state measurement on qubits 1 and 8 and announces the measurement outcomes publicly. Finally, Bob, Charlie, and David perform a Bell-state measurement on their own qubit pairs in turn [see Fig.1 (c)]. The measurement results are their own pieces of secret.

---

[*] Corresponding author. Email address: lins95@fjnu.edu.cn

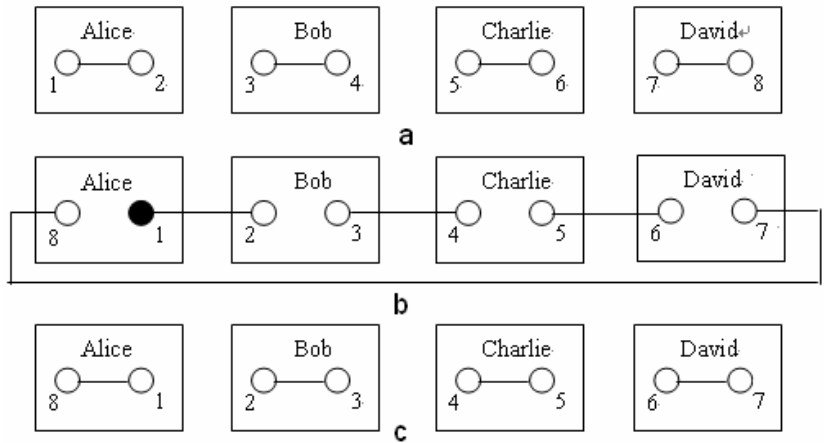

FIG.1.The Zhang-Man protocol for four-party quantum secret sharing. Each circle represents a qubit and the solid one denotes a qubit on which a unitary operation will be, respectively. The line between qubits indicates their entanglement.

During the process of reconstructing the secret, Bob, Charlie, and David cooperate to deduce the local unitary operation that Alice performed on qubit 1 according to their measurement outcomes and Alice's announced message. Then they can attain the secret.

In Ref. [6], the author claimed this MQSS protocol is secure. However, if Bob and David, the unauthorized agent group, are dishonest, they can steal Alice's secret without the help of Charlie by using the following strategy. In the detecting mode, Bob and David act according to the legal process. But in the message mode, Bob sends the qubit 2 to David and David sends the qubit 6 to Bob. After that, Bob performs a Bell-state measurement on the qubits 3 and 6 instead of that on 2 and 3. Similarly, David makes the same measurement on the qubits 2 and 7 instead of that on 6 and 7. By this way, as depicted in Fig. 2, Bob performed an entanglement swapping with Charlie while David with Alice. Therefore, David can easily deduce Alice's operation on the qubit 1 (i.e., the secret of Alice), and Bob can obtain Charlie's measurement result (i.e., the piece of secret of Charlie) according to the rule of entanglement swapping. For instance, if Alice's measurement result is $|\Phi^+\rangle_{18}$ and David's is $|\Psi^+\rangle_{27}$, David knows Alice's unitary operation is $u_4$, that is, the secret bits are "11", On the other hand, if Bob's measurement outcome is $|\Phi^-\rangle_{36}$, he knows Charlie's result is $|\Phi^-\rangle_{45}$.

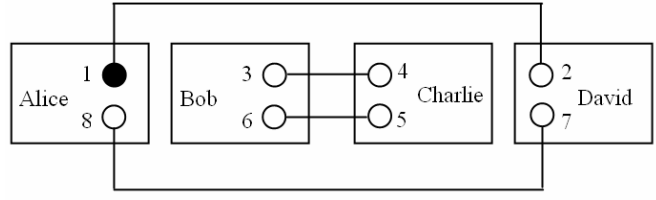

FIG.2. A two special dishonest parties cooperate attack on Zhang-Man's protocol.

Since this attack only happens in the message mode, Alice can't detect whether Bob makes Bell-state measurement according to the protocol. So this attack would not introduce any error. A desirable question is whether it can be detected if Alice does an additional detection as other QKD protocols generally do, that is, Alice requires all her agents announce some sampled key bits and compares them after the key distribution was finished (but before Alice's declaration of her outcomes). The answer is negative. Let $|\varphi\rangle_{36}$ and $|\varphi\rangle_{27}$ are the measurement results of Bob and David, respectively. In the additional detection, Bob and David can escape successfully by announcing $U|\varphi\rangle_{36}$ and $U|\varphi\rangle_{27}$, where $U \in \{u_1, u_2, u_3, u_4\}$. Such announcements will not introduce any error. We consider the same scenarios as that of the above example. Hence, Charlie's results are $|\Phi^-\rangle_{45}$. Bob announce a fake information, which his results are $|\Phi^-\rangle_{23}$. By the same way, David publish that his results are $|\Psi^+\rangle_{67}$. Because

$$|\Phi^-\rangle_{12} \otimes |\Psi^-\rangle_{78} = \frac{1}{2}(|\Phi^-\rangle_{18} \otimes |\Psi^-\rangle_{27} + |\Phi^+\rangle_{18} \otimes |\Psi^+\rangle_{27} + |\Psi^-\rangle_{18} \otimes |\Phi^-\rangle_{27} + |\Psi^+\rangle_{18} \otimes |\Phi^+\rangle_{27}) \quad (1)$$

$$|\Psi^+\rangle_{27} \otimes |\Psi^-\rangle_{34} = \frac{1}{2}(|\Phi^-\rangle_{23} \otimes |\Phi^+\rangle_{47} + |\Phi^+\rangle_{23} \otimes |\Phi^-\rangle_{47} + |\Psi^+\rangle_{23} \otimes |\Psi^-\rangle_{47} + |\Psi^-\rangle_{23} \otimes |\Psi^+\rangle_{47}) \quad (2)$$

$$|\Phi^+\rangle_{47} \otimes |\Psi^-\rangle_{56} = \frac{1}{2}(|\Phi^-\rangle_{45} \otimes |\Psi^+\rangle_{67} + |\Phi^+\rangle_{45} \otimes |\Psi^-\rangle_{67} + |\Psi^+\rangle_{45} \otimes |\Phi^-\rangle_{67} + |\Psi^-\rangle_{45} \otimes |\Phi^+\rangle_{67}) \quad (3)$$

The results of Bob, Charlie, and David are match to Alice's previous state of the EPR pair (1, 2) and her measurement outcomes on the EPR pair (1, 8). Therefore, no error was introduced, i.e., Alice cannot detect eavesdropping even if she performs a new detection at the end of the process of Zhang-Man protocol.

The above eavesdropping strategy is easy to be generalized to a multiparty case. We assume that Alice splits her secret into *n* parts and distributes it to *n* agents, $Bob_1, ..., Bob_n$. According to the attack strategy described above, $Bob_1$ and $Bob_n$ can cooperate and eavesdrop the secret without being detected. More generally, if $Bob_i$ and $Bob_j$ (*j*>*i*) attack this protocol collectively, they can make the pieces of secret of the other agents between them i.e. $Bob_k$ (*i*<*k*<*j*) unnecessary in the process of reconstructing the secret, that is, the remainders can attain Alice's secret without the help of these agents. Nevertheless, in Zhang-Man protocol, it requests that no subset of agents is sufficient to attain the secret. Hence, Zhang-Man protocol may be insecure if there are two or more dishonest parties.

To improve the security of Zhang-Man protocol, we make a modification on it so that the improved protocol can stand against this kind of attack. For convenience，we also discuss a four-party QSS protocol as illustrated in Fig.3.

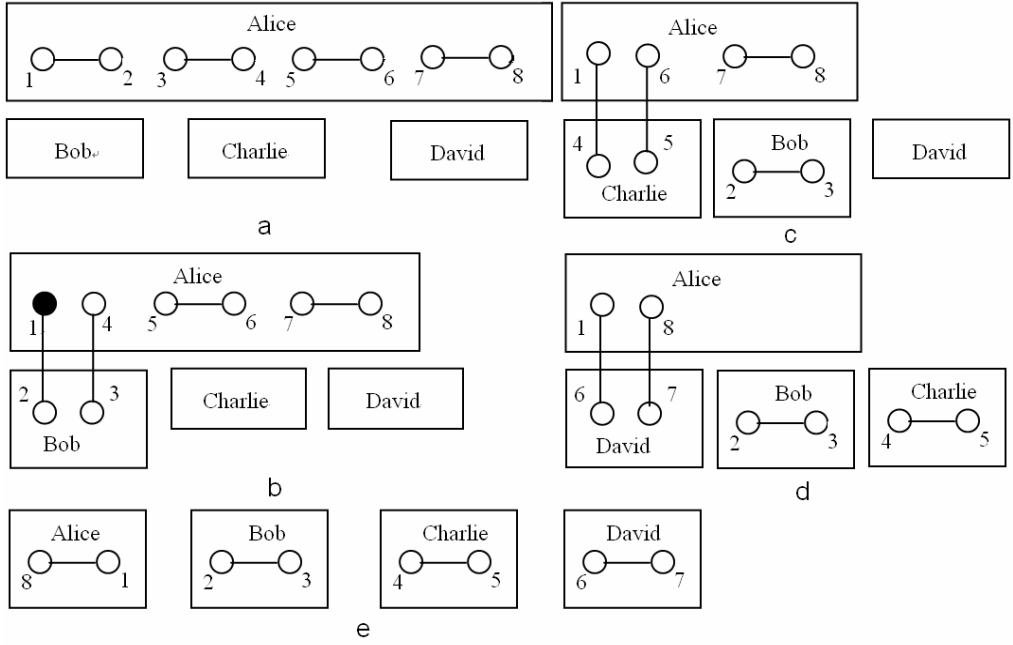

FIG.3. A improved MQSS protocol.

(1) Alice prepares four EPR pairs $|\Psi^-\rangle_{12}, |\Psi^-\rangle_{34}, |\Psi^-\rangle_{56}, |\Psi^-\rangle_{78}$ [see Fig.3 (a)].

(2) Alice sends qubits 2 and 3 to Bob. After Bob received these qubits [see Fig.3 (b)], Alice randomly chooses the following two procedures: a) The detecting mode. Alice chooses at random whether to measure qubit 1 in the rectilinear basis $\{|0\rangle, |1\rangle\}$ or the diagonal basis $\{|+\rangle, |-\rangle\}$. After taking the same measurement on qubit 2, Bob tells Alice his measurement outcome through the classical channel. Alice compares their measurement results and detects eavesdropping. b) The message mode. Alice chooses randomly one of the four local unitary operations, $\{u_1, u_2, u_3, u_4\}$, which represent Alice's secret, and performs this operation on the qubit 1, which encodes the two secret bits on the Bell state. Then Alice measures the Bell operator on qubits 1 and 4. Bob makes a Bell-state measurement on qubits 2 and 3. The results of this measurement define his piece of secret.

(3) Alice sends qubits 4 and 5 to Charlie. After Charlie received these qubits[see Fig.3 (c)], Alice randomly chooses either of the following procedures: a) The detecting mode. The concrete process is the same as that in step (2). b) The message mode. Alice measures the Bell operator on qubits 1 and 6. Charlie makes a Bell-state measurement on qubits 4 and 5. The results of this measurement define his piece of secret.

(4) In terms of the same method, Alice sends qubits 6 and 7 to David and he can attain his piece of secret by Bell-state measurement.

(5) Alice announces the measurement outcomes on qubits 1 and 8 publicly.

It can be simply verified that Bob, Charlie, and David can reconstruct the secret according to the rule of entangled swapping if they collaborate. For example, if the results of Alice, Bob, Charlie, and David are $|\Phi^+\rangle_{18}$, $|\Phi^-\rangle_{23}$, $|\Phi^-\rangle_{45}$ and $|\Psi^-\rangle_{67}$ respectively, since the initial states of the qubits 7 and 8 is $|\Psi^-\rangle_{78}$, they can know that the EPR pair (1, 6) have been projected to $|\Phi^+\rangle_{16}$ in terms of Alice's and David's results. By the same way, they can deduce the state $|\Psi^+\rangle_{14}$ of the EPR pair (1, 4) and $|\Phi^+\rangle_{12}$ of the EPR pair (1, 2). Finally, they can determine that the secret is "10". It should be emphasized that the users need not restart the protocol if Alice chooses the detecting mode in step (3) and (4), since Alice can prepare a new EPR pair in the same Bell state to resume the protocol.

When Alice measures the Bell operation on qubits 1 and 4, the secret is split into two parts, which are the measurement outcomes of the EPR pair (2, 3) and (1, 4). So if Bob and David want to attain the secret, they should know the Bell state of the qubits 1 and 4. Because the EPR pair (1, 4) is still in Alice's site, Bob and David cannot obtain any information about the secret even if Bob sends qubit 2 to David. Then Alice split it into two parts by entanglement swapping, which are the Bell states of the EPR pair (1, 6) and (4, 5). Since the channel between Alice and Charlie is secure, only Charlie knows the measurement results on the EPR pair (4, 5) and anyone cannot attain it without being detected. Hence Bob and David cannot recover the secret without the assistance of Charlie. It means that the improved protocol can stand against the attack presented in this paper. When Alice detects the security of the channel between Alice and her agent, only the agent's outcomes are required without the cooperation of any other agent. So the improved protocol is also secure against the attack proposed in Ref. [8].

In summary, we have proposed a special attack strategy to Zhang-Man protocol [6], in which unauthorized agent group can collaborate and eavesdrop the secret without being detected. Furthermore, an improvement on Zhang-Man protocol, which makes it secure against this kind of attack, is presented. Finally, we make a simple security analyses on the improved protocol.

**Acknowledgements**

This work is supported by the National Natural Science Foundation of China under Grant No. 60373059, the Major Research Plan of the National Natural Science Foundation of China under Grant No. 90604023, the National Research Foundation for the Doctoral Program of Higher Education of China under Grant No. 20040013007, the National Laboratory for Modern Communications Science Foundation of China, Grants No. 9140C1101010601, the Graduate Students Innovation Foundation of BUPT and the ISN Open Foundation.

**Reference**
[1] M. Hillery, V. Bužek and A. Berthiaume, Phys. Rev. A 59 (1999) 1829.
[2] A. Karlsson, M. Koashi and N. Imoto, Phys. Rev. A 59 (1999) 162.
[3] R. Cleve, D. Gottesman and H.K. Lo, Phys. Rev. Lett. 83 (1999) 648.
[4] D. Gottesman, Phys. Rev. A 61 (2000) 042311.
[5] G.P. Guo and G.C. Guo, Phys. Lett. A 310 (2003) 247.
[6] Z.J. Zhang and Z. X. Man, Phys. Rev. A **72**, 022303(2005)

[7] F. Gao, Q.Y. Wen and F.C. Zhu, Physics Letters A 360 (2007) 748–750

[8] F.G. Deng, X.H. Li, P. Chen, C.Y. Li and H.Y. Zhou, e-print quant-ph/0604060.